\documentclass[aps,pre,twocolumn,groupedaddress,longbibliography]{revtex4-1}
\usepackage[dvipsnames]{xcolor}
\usepackage{graphicx,amsmath,amssymb,color}
\begin{document}

\author{Robert S. Hoy}
\email{rshoy@usf.edu}
\author{Kevin A. Interiano-Alberto}
\affiliation{Department of Physics, University of South Florida, Tampa, FL 33620 USA}
\date{\today}

\title{Efficient $d$-dimensional molecular dynamics { simulations} for studies of the glass-jamming transition}

\begin{abstract}
We develop an algorithm suitable for parallel molecular dynamics simulations in $d$ spatial dimensions and describe its implementation in C++.
All routines work in arbitrary $d$; the maximum simulated $d$ is limited only by available computing resources.
These routines include several that are particularly useful for studies of the glass/jamming transition, such as SWAP Monte Carlo and FIRE  energy minimization.
Scaling of simulation runtimes with the number of particles $N$ and number of simulation threads $n_{\rm threads}$ is comparable to popular MD codes such as LAMMPS.
The efficient parallel implementation allows simulation of systems that are much larger than those employed in previous high-dimensional glass-transition studies.
As a demonstration of the code's capabilities, we show that  supercooled $d = 6$ liquids can possess dynamics that are substantially more heterogeneous and experience a breakdown of the Stokes-Einstein relation that is substantially stronger than previously reported, owing at least in part to the much smaller system sizes employed in earlier simulations.
\end{abstract}
\maketitle

\section{Introduction}
\label{sec:intro}

Molecular dynamics (MD) simulations have been an essential part of statistical physicists' toolbox for over 50 years \cite{hoover68,weeks71,allen87,frenkel02}.
Numerous open-source, highly-optimized multipurpose parallel MD simulation packages are now available \cite{lammps,gromacs,namd,amber,hoomd,rumd}.
These packages are designed to simulate systems embedded in the physically relevant spatial dimensions $2 \leq d \leq 3$; they cannot be used for $d > 3$ simulations without extensive modifications.
Accordingly, they employ parallelization algorithms that are efficient in low $d$, \textit{e.g.}\ spatial domain decomposition allowing simulation of multibillion-atom systems on distributed-memory supercomputers \cite{plimpton95,glosli07}.
These algorithms, however, rapidly become less efficient as $d$ increases.
Developing publicly available codes that allow efficient parallel MD simulations of higher-dimensional systems to be performed may be a key step towards answering several open questions in physics, particularly questions related to supercooled liquids and the glass/jamming transition.
In this paper, we describe the first such code (\texttt{hdMD}) \cite{pyCUDAsuss}.
Then we use it to show that the dynamics of supercooled $d = 6$ liquids at densities above the RFOT dynamical glass transition density $\phi_d$ \cite{kirkpatrick89} can be substantially more heterogeneous than previously reported.

Early simulations of high-dimensional supercooled liquids employing systems of $N = 10^3-10^4$ particles indicated that crystallization is strongly suppressed as $d$ increases beyond $3$ \cite{skoge06,bruning09,vanMeel09b,charbonneau11}.
Recent work, however, has shown that (i) crystallization instabilities in glassforming mixtures often appear only when larger $N$ \cite{ingebrigtsen19} or SWAP Monte Carlo equilibration \cite{ninarello17} are employed, and (ii) hard-sphere crystals are thermodynamically stable for a wide range of packing fractions $\phi < \phi_d$ for all $d \leq 10$ \cite{charbonneau21}.
Both results suggest that the conclusions of many of the early $d > 3$ studies need to be reexamined using larger systems.
Moreover, the $N$ typically employed in recent simulations of higher-$d$ systems have remained small (e.g.\ $N < 10^4$ in Refs.\ \cite{berthier19c,berthier20,morse21}), precluding robust investigation of how any static or dynamic length scales that grow substantially as $\phi_d$ is approached \cite{berthier05,kob97,donati98,starr13,karmakar14} depend on $d$.

For example, simulation studies suggesting that dynamical heterogeneity weakens as $d$ increases \cite{charbonneau12,charbonneau13b,adhikari21} can be challenged on the grounds that heterogeneous dynamics in these studies are suppressed not (or not primarily) by the increase in spatial dimension, but instead by their use of periodic boundary conditions with simulation cell side lengths $L$ that decrease rapidly with increasing $d$.
This is a reasonable concern given the tendency of the least mobile particles in glassforming liquids to remain near their initial positions after time intervals over which the most mobile particles have hopped by several times their diameter \cite{weeks00,chaudhuri07} and the fact that $L$ drops as low as $\sim 2$ particle diameters for the highest studied $d$ \cite{charbonneau12,charbonneau13b,adhikari21}.
Resolving this issue requires performing analogous simulations with larger $L$, and completing such simulations in a reasonable amount of time requires an efficient parallel implementation of high-dimensional MD.

Simulating supercooled liquids in high $d$ presents several challenges.
One of the most difficult involves neighbor-list construction.
If one wishes to simulate systems with a given number of particles $N$, the cell side length $L \sim N^{1/d}$, and thus the number of linked (fixed-size) subcells along each axis of the simulation cell is $n_{\rm cell} \sim N^{1/d}$. 
The total number of subcells is $\mathcal{N}_{\rm sc} = n_{\rm cell}^d$, and the number of these which must be searched over for each atom during a typical Verlet list (VL) build is $3^d$.
The total number of atoms in these neighboring cells is  $N_{nc} = (3^d/\mathcal{N}_{\rm sc})N \equiv (3/n_{\rm cell})^d N \sim 3^d$, so the total number of pair distances which must be evaluated to rebuild all atoms' Verlet lists is $\mathcal{N}_{\rm Vl} \sim 3^d N$.  
In other words, the effort required to maintain the neighbor lists increases exponentially with increasing $d$ because building each atom's VL requires searching over a exponentially increasing fraction of the system's volume.

Avoiding strong finite-size effects requires simulating systems of a minimum spatial extent $\mathcal{L}$; typically $\mathcal{L} \gtrsim 10$ particle diameters. 
If $\mathcal{L}$ is taken to be $L$, the minimum number of particles simulations must include to avoid such affects is $N_{\rm min} \sim L^d$.
While characteristic packing fractions $\phi$ scale across dimensions roughly as  $2^{-d}$ or $d2^{-d}$ \cite{charbonneau17,parisi10}, hypersphere volumes decrease faster than this, so the corresponding particle number densities $\rho$ increase with increasing $d$.
For example, the dynamical-glass-transition number densities for hard spheres, $\rho_d \simeq 1.78 - .5d + .095d^2$ \cite{charbonneau12}, nearly triple as $d$ increases from 3 to 7.
Taking $N_{\rm min}  = \rho_d L^d$, the computational effort to build the VL for a $d$-dimensional supercooled liquid scales roughly as $\rho_d (3L)^d$.

This rapidly becomes prohibitive as $d$ increases, but it can be overcome for intermediate $d$ using efficient parallel algorithms.
In this paper, we describe our newly developed \texttt{hdMD} code, which employs several such algorithms and is capable of simulating much larger high-$d$ systems than have been studied previously.
\texttt{hdMD} includes several routines commonly employed in studies of the glass/jamming transition \cite{debenedetti01,liu10,berthier11}, 
such as SWAP Monte Carlo \cite{grigera01,ninarello17}, FIRE  energy minimization \cite{bitzek06,guenole20}, and calculation of the overlap parameter $f_{\rm ov}(t)$ and van Hove correlation function $G_s(r,t)$.
Here we use it to simulate $3 \leq d \leq 6$ supercooled liquids of up to $10^7$ particles over short timescales, and $10^5$ particles over the much longer timescales typically employed in modern glass-transition studies \cite{ninarello17}.
The latter simulations show that strongly heterogenous dynamics [as indicated by large non-Gaussian parameters] can persist to very long times in deeply supercooled liquids for $d$ up to at least $6$.
By contrasting results for $N = 10^5$ to those for $N = 5000$ liquids at the same  $\phi$ and  $T$, we provide initial evidence that some of the conclusions of Refs.\  \cite{charbonneau12,charbonneau13b,adhikari21} -- at least in terms of their quantitative details -- would have been different had these studies considered larger systems.

The outline of the rest of this paper is as follows.
Section \ref{sec:alg} describes  \texttt{hdMD}'s algorithmic implementation.
Section \ref{sec:bench} presents performance benchmarks for simulations at packing fractions $\phi \simeq \phi_d$ for $3\leq d \leq 6$, focusing on the code's parallel efficiency and the scaling of simulation runtimes with $N$ and $d$.
Section \ref{sec:gd36} includes an original analyisis of the dynamics of supercooled $d = 6$ liquids
and serves as a demonstration of \texttt{hdMD}'s suitability for state-of-the-art glass-transition-related studies.
Finally, in Section \ref{sec:dc}, we discuss our results and conclude.
Readers more interested in the code's performance than its algorithmic implementation are urged to skip to Sec.\ \ref{sec:bench}, while readers primarily interested in our supercooled-liquid physics results are urged to skip to Sec.\ \ref{sec:gd36}.

\section{Algorithmic Implementation}
\label{sec:alg}

Most previous simulations of liquids in $d > 3$ have employed hard spheres. 
This is sensible given that hard-core interactions dominate liquids' structure \cite{weeks71} and polydisperse hard-sphere liquids are excellent glassformers \cite{ninarello17}.
However, hard-particle techniques are inherently limited in the range of physical phenomena they can capture.
The lack of finite interparticle forces makes it difficult for them to accurately model the collective rearrangements which increasingly dominate liquids' relaxation mechanisms as the glass transition is approached \cite{kob97,donati98}.
They cannot, for example, capture the long-range elastically-mediated dynamical facilitation that has recently been shown to arise below the mode-coupling temperature $T_{\rm MCT}$ \cite{chacko21}.
While attractive forces are well known to exert a strong and nonperturbative influence on 3D glassforming liquids' dynamics  \cite{berthier09c} and to increase their tendency to crystallize \cite{toxvaerd21}, the variation of these effects with increasing $d$ remains unexplored and cannot be explored using hard-particle models.
Therefore we employ a soft-particle MD approach that can treat both repulsive and attractive forces.

Parallelizing a MD simulation requires devising a method of distributing the computation across its $n_{\rm threads}$ concurrent threads.
Modern MD packages, most of which are optimized for large-$n_{\rm threads}$ simulations on distributed-memory machines  \cite{lammps,gromacs,namd,amber,hoomd,rumd}, typically parallelize via spatial domain decomposition of simulation cells \cite{plimpton95}.
In this scheme, different CPU cores ``own'' geometrically distinct regions, and information that needs to be passed across the boundaries of these regions is typically passed via intercore communication, usually implemented using MPICH.
For example, a typical $n_{\rm threads} = 8$ simulation employing a cubic simulation cell divides it into $2\times 2 \times 2$ domains, each of which is a cube with $n_{\rm cell}/2$ subcells along each edge.
The intercore communication that is necessary for force evaluation and Verlet list (VL)-building  is only required between subcells on the surfaces of these domains \cite{plimpton95}.

Codes that operate this way work very well for $d \leq 3$.
For example, simulations of million-atom liquids have been shown to exhibit nearly-optimal scaling (runtimes $\sim n_{\rm threads}^{-1}$) for $n_{\rm threads}$ up to $\sim 10^2$ \cite{plimpton12}.
As $d$ increases, however, the spatial-domain-decomposition method loses its effectiveness for reasons comparable to those outlined above for VL-building.
Therefore we respectively employ per-atom, per-atom, and per-subcell parallelization for force evaluation, time integration of equations of motion, and VL-building.
In this scheme, each thread is responsible for $N/n_{\rm threads}$ atoms and $\mathcal{N}_{\rm sc}/n_{\rm threads}$ subcells rather than a spatial domain, and intercore communication is avoided entirely by using OpenMP rather than MPICH to parallelize the code.
While the latter choice limits our code to shared-memory (as opposed to distributed-memory) machines, it makes it far more efficient.

\begin{table}[h]
\caption{Principal routines (C++ functions) included in \texttt{hdMD}. 
Additional standard functions involving I/O and memory allocation are present in the code but are not listed here.}
\begin{ruledtabular}
\begin{tabular}{ll}
Name & Purpose \\
\texttt{main} & Control program operation \\
\texttt{leapfrog} & Integrate Newton's EOM \\
\texttt{getforce} & Calculate force on atom $i$ \\
\texttt{Berendsen} & Berendsen thermo/barostat \\
\texttt{getthermo} & Calculate thermodynamic quantities \\
\texttt{getvanHove} & Calculate van Hove corr.\ function \\
\texttt{initRMSE} & Initialize $\{ \vec{r}_i \}$, $\{ m_i \}$, $\{ \epsilon_i \}$, $\{ \sigma_i \}$  \\ 
\texttt{setupcell} & Initialize simulation cell \\ 
\texttt{needsrebuild} & Check if VLs need to be rebuilt \\
\texttt{buildneighborlist} & Build/rebuild the VLs \\
\texttt{FIRE} & Control FIRE energy minimization \\
\texttt{FIREintegrate} & Integrate FIRE EOM \\
\texttt{steepest} &  Steepest-descent energy minimization \\
\texttt{buildswapNLs} & Generate SWAP-attempt VLs \\
\texttt{swapmove} & Attempt SWAP moves \\
\texttt{writerestart} & Write restart files to disk \\
\texttt{writecoords} & Write $\{ \vec{r}_i \}$ and image flags to disk
\end{tabular}
\end{ruledtabular}
\label{tab:fnlist}
\end{table}

Table \ref{tab:fnlist} lists \texttt{hdMD}'s principal routines.
All of these routines work in arbitrary $d$; the maximum simulated $d$ is limited only by available computing resources.
Our implementation of these routines will be described in detail in the following sections.

\subsection{Particle model and initial state generation}
\label{subsec:model}

Here, for simplicity, we present results for a single hard-sphere-like pair potential, the truncated and shifted Morse potential given by
\footnotesize
\begin{equation}
U_a(\epsilon_{ij}, \sigma_{ij},r_{ij}) = 
\epsilon_{ij} \bigg{ [ } \exp[-2a (r_{ij} - \sigma_{ij})] - 2\exp[-a (r_{ij} - \sigma_{ij})]  + 1 \bigg{ ] } 
\label{eq:morsepot}
\end{equation}
\normalsize
for $r_{ij} \leq \sigma_{ij}$ and zero for $r_{ij} \geq \sigma_{ij}$, where $r_{ij}$ is the center-to-center distance between particles $i$ and $j$. 
We use the standard Lorentz-Berthelot MD mixing rules for particle diameters and force prefactors, i.e.\ $\sigma_{ij} = (\sigma_i + \sigma_j)/2$ and $\epsilon_{ij} = \sqrt{\epsilon_i \epsilon_j }$.
Longer-range interactions including attractive terms can be implemented by adjusting \texttt{main()}'s ``cut'' parameter.
The Morse parameter $a$ can be adjusted by editing a single-line in \texttt{getforce()}; all results presented below are for $a = 30\tilde{\sigma}^{-1}$. 
More generally, \texttt{hDMD} can be easily generalized to arbitrary radial force laws and mixing rules (incorporating, e.g., nonadditivity \cite{ninarello17})  by editing a few lines of code in the \texttt{getforce()}, \texttt{thermo()}, and \texttt{swapmove()} routines.
In particular, switching the stock code from the Morse to the Mie potential $U_n(\epsilon_{ij}, \sigma_{ij},r_{ij}) = \epsilon_{ij} [ ( \frac{ \sigma_{ij} }{r} )^{2n} - 2 ( \frac{ \sigma_{ij} }{r} )^{n}]$, which reduces to the widely employed Lennard-Jones/WCA potential for $n=6$, requires only commenting/uncommenting-out a few lines of code in these routines.

The volume of a $d$-dimensional spherical particle with diameter $\sigma$ is
\begin{equation}
v(d,\sigma) =  \displaystyle\frac{\pi^{d/2} \sigma^d }{2^d \Gamma(1 + d/2)},
\label{eq:vofsigma}
\end{equation}
where $\Gamma$ is the gamma function.
For all simulations discussed in Section \ref{sec:bench}, we set particles' mass density $\rho_m = [v(d,\sigma)\sigma^{-d}]^{-1} = \pi^{-d/2} 2^d \Gamma(1 + d/2)$ so particles have mass $m_i = \sigma_i^d$ and typical particles with $\sigma = 1$ have unit mass.
Correspondingly, we set $\epsilon_i = \sigma_i^d$ so repulsive interactions also scale with particle volumes \cite{footepsscaling}.
During MD simulation,  Newton's equation of motion are integrated using the standard leapfrog algorithm with a timestep $dt = 6\tau/(125a) = .0016\tau$, where $\tau = \sqrt{\tilde{m}\tilde{\sigma}^2/\tilde{\epsilon}}$ is the unit of time.
Here $\tilde{m}$, $\tilde{\sigma}$, and $\tilde{\epsilon}$ are the mass, diameter and force prefactor for typical particles: $\tilde{m} = \tilde{\sigma} = \tilde{\epsilon} = 1$ in dimensionless units.

Polydispersity is one of the most important factors controlling glassforming ability in molecular simulations \cite{williams01}.
We use the particle-diameter distribution 
\begin{equation}
P(d, \mathcal{R}, \sigma) = \Bigg{ \{ }
\begin{array}{ccc}
\displaystyle\frac{ (d-1)\sigma^{-d} }{ \mathcal{R}^{\frac{d-1}{2d}} - \mathcal{R}^{-\frac{d-1}{2d}} } & , & \mathcal{R}^{-\frac{1}{2d}} \leq \sigma \leq  \mathcal{R}^{\frac{1}{2d}} \\
\\
0 & , & \sigma < \mathcal{R}^{-\frac{1}{2d}} \ \rm{or}\ \sigma > \mathcal{R}^{\frac{1}{2d}}
\end{array},
\label{eq:sigmadist}
\end{equation}
where $\mathcal{R} = (\sigma_{\rm max}/\sigma_{\rm min})^d$ is the ratio of maximum to minimum particle volumes.
The total volume $\mathcal{V}(\sigma) = NP(d, \mathcal{R}, \sigma)  v(d,\sigma)$ occupied by particles of diameter $\sigma$ is $\sigma$-independent over the entire range $\mathcal{R}^{-\frac{1}{2d}} \leq \sigma \leq  \mathcal{R}^{\frac{1}{2d}}$; this choice apparently optimizes glass-formability for a wide range of force laws \cite{ninarello17}.
Preventing crystallization also requires a sufficiently large polydispersity index $\Delta = [\langle \sigma^2 \rangle - \langle \sigma \rangle^2]^{1/2}/ \langle \sigma \rangle$, with the minimum $\Delta$ decreasing with increasing $d$ \cite{ninarello17}.
Here we choose $\mathcal{R} = d$, which gives $\Delta = 0.107, 0.099, 0.091, 0.083$ in $d = 3-6$.
We find that these parameter choices are sufficient to prevent both crystallization and phase separation in all systems discussed below.

To generate $N$-particle initial states with packing fraction $\phi$, we first define the particle diameters $\{ \sigma_i \}$ by randomly sampling the distribution $P(d, \mathcal{R}, \sigma)$, using the function \texttt{initRMSE()}.
This function also sets the particle masses $\{ m_i \}$ and force prefactors $\{ \epsilon_i \}$.
The total volume occupied by these particles is
\begin{equation}
V_{\rm part} = \displaystyle\sum_{i = 1}^N v(d,\sigma_i);
\end{equation}
thus the total simulation cell volume must be $V = V_{\rm part}/\phi$.
We employ hypercubic simulation cells of this volume (and side length $L = V^{1/d}$), centered at the origin; \texttt{initRMSE()} assigns all particles random initial positions $\{ \vec{r}_i \}$ within these cells.
After initializing the linked-subcell data structures using \texttt{setupsupercell()}, we populate the subcells and initialize all particles' Verlet lists using \texttt{buildneighborlist()}.
These two routines will be described in detail in Section \ref{subsec:NLbuilds}.

The randomly generated positions lead to severe particle overlap that must be reduced before the simulation can begin. 
We accomplish this ``pushoff'' using a partial FIRE minimization (Section  \ref{subsec:integration}).
During this minimization and throughout the rest of the simulation, periodic boundary conditions are applied along all $d$ directions.
After the pushoff is complete, we give particles random initial velocities corresponding to the desired target temperature $T_{\rm targ}$.
Alternatively, the abovementioned initial-state-generation-and-pushoff procedure may be skipped by reading initial states (e.g.\ restart files written during a previous simulation) in from an ASCII file.
This is also handled using \texttt{initRMSE()}, with the relevant restart file names passed to \texttt{hdMD} as command-line arguments as discussed below.
Then the linked-subcell data structures are initialized, the subcells are populated and all particles' VLs are initialized as outlined above. 
Once either of these options is completed, the MD simulation begins.

\subsection{Program control, force evaluation and  integration of equations of motion}
\label{subsec:integration}

In this section, we outline \texttt{hdMD}'s usage and large-scale structure.
Entering the command \texttt{./hdMD $d$ $N$ $\phi$ $T$ $n_{\rm steps}$ $n_{\rm threads}$ $p$ resfilename rinitfilename $n_{\rm msdstart}$}
starts a $d$-dimensional MD simulation with $N$ particles at packing fraction $\phi$ and target temperature $T$.  
The fifth and sixth command-line parameters specify the number of MD timesteps and the number of parallel OpenMP threads.
$p$ is the SWAP attempt fraction, i.e.\ the fraction of particles for which swaps are attempted every time \texttt{swapmove()} gets called.
 \texttt{resfilename} is the name of the file containing the initial particle $\{ m_i \}$, $\{ \sigma_i \}$, $\{ \epsilon_i \}$, positions, velocities, and image flags.
 Such a file is unnecessary (and \texttt{resfilename} should be ``null'') if these are to be generated as outlined above.
Finally, \texttt{rinitfilename} is the name of the file containing an \textit{earlier} set of particle positions and image flags to be used as the reference configuration for calculations of particles' mean-squared displacement, etc.\ as outlined in Section \ref{subsec:auxfcns}.
Such a file is unnecessary and  \texttt{rinitfilename} should  be ``null'' if the initial particle positions are to be used for this  reference configuration.
In addition to the abovementioned command-line parameters, a number of other runtime parameters are defined in  \texttt{hdMD}'s \texttt{main.cpp} file and summarized in Table \ref{tab:parlist}.

Once a simulation is running, it dumps thermodynamic data every $n_{\rm thermo}$ MD timesteps to a file named \texttt{thermodata.d$d$.N$N$.phi$\phi$.p$p$}, energy-minimization data every $n_{\rm min}$ timesteps to a file named \texttt{mindata.d$d$.N$N$.phi$\phi$.p$p$} (if \texttt{minflag} = true), restart files including all particles $\{ m_i \}$, $\{ \sigma_i \}$, $\{ \epsilon_i \}$, positions, velocities, and image flags every $n_{\rm restart}$ timsteps to files named \texttt{restart.d$d$.N$N$.phi$\phi$.p$p$.step$step$}, van Hove correlation function data  every $n_{\rm vhc}$ timesteps  to a file named \texttt{vanHoveCorr.d$d$.N$N$.phi$\phi$.p$p$} (if \texttt{vhcflag} = true), and the particle positions and image flags at step $n_{\rm msdstart}$ to a file named \texttt{Rinit.d$d$.N$N$.phi$\phi$.p$p$.step$step$} (and also every $n_{\rm dump}$ timesteps to a file named \texttt{coords.d$d$.N$N$.phi$\phi$.p$p$}).
Italicized quantities indicate the numerical values of the various parameters.

\begin{table}[h]
\caption{Principal user-adjustable parameters defined in \texttt{hdMD}.  Each of these may be adjusted by editing one line of code in \texttt{main()}.  Alternatively, they may be straightforwardly converted to command-line parameters.}
\begin{ruledtabular}
\begin{tabular}{ll}
Name & Definition \\
$n_{\rm thermo}$ & Steps between \texttt{getthermo()} calls \\
$n_{\rm vhc}$ & Steps between \texttt{getvanHove()} calls \\
$n_{\rm baro}$ & Steps between \texttt{Berendsen()} calls \\
$n_{\rm restart}$ & Steps between \texttt{writerestart()} calls \\
$n_{\rm dump}$ & Steps between \texttt{writecoords()} calls \\
$n_{\rm bet}$ & Steps between \texttt{needsrebuild()} calls \\
$n_{\rm min}$ & Steps between \texttt{FIRE()} calls \\
$n_{\rm swap}$ & Steps between \texttt{swapmove()} calls \\
$n_{\rm msdstart}$ & Steps before MSD calculation begins \\
$n_{\rm chunk}$ & Chunk size for integration of EOM \\
$T_{\rm targ},\ P_{\rm targ}$ & Target temperature and pressure \\
$\tau_{\rm temp},\ \tau_{\rm press}$ & Temperature and pressure damping times \\
$E_{\rm thres},\ F_{\rm thres}$ &  Convergence criteria for \texttt{FIRE()} \\ 
maxiter & Maximum number of FIRE iterations \\
$\Delta_{\rm max}$ & Maximum particle displacement/timestep \\
baroflag & True (false) for NPT  (NVT) simulations \\
minflag & True (false) if performing (not performing) \\
& periodic energy minimization \\
vhcflag & True (false) if performing (not performing) \\
& van Hove correlation function calculations \\
\end{tabular}
\end{ruledtabular}
\label{tab:parlist}
\end{table}

After initial states are prepared using either of the two methods outlined above, \texttt{hdMD}'s main loop (i.e.\ time integration for $n_{\rm steps}$ MD timesteps) begins.  
Before describing the structure of this main loop, however, we discuss some implementation details particular to \texttt{hdMD} that we found optimize its performance.
First, rather then employing three 2D arrays of format $\texttt{r[N][d]}$,  $\texttt{v[N][d]}$, and  $\texttt{f[N][d]}$ for the positions, velocities and forces as is done in (e.g.)\ LAMMPS \cite{lammps}, we store all three in a single 1D array (\texttt{rvf}) of length $3dN$.
Second, for all parallel for loops of length $\sim dN$, we employ the OpenMP scheduling directive \texttt{schedule (static,$n_{\rm chunk}$)} with $n_{\rm chunk} = 100$.
Both of these choices speed up the code by improving its cache locality; see Section \ref{subsec:scal}.

\texttt{hdMD}'s main loop is structured as follows:
\begin{itemize}
\item \texttt{needsrebuild()} is called once every $n_{\rm bet}$ timesteps to determine whether particles' VLs need to be rebuilt  (Section \ref{subsec:NLbuilds}).
\item On timestep $n_{\rm msdstart}$, the positions \texttt{rinit} that will be used as the reference state for followup calculcations of particles' mean-squared displacement, van Hove correlation function, etc.\ (Section \ref{subsec:auxfcns}) are stored in memory and written to \texttt{rinitfile}.
\item \texttt{getthermo()} is called once every $n_{\rm thermo}$ timesteps.
\item If \texttt{vhcflag} is set to  \texttt{true} and  and  $\texttt{step} \geq n_{\rm msdstart}$, \texttt{getvanHove()} is called once every $n_{\rm vhc}$ timesteps.
\item \texttt{writerestart()} is called once every  $n_{\rm restart}$ timesteps.
\item If \texttt{minflag} is set to  \texttt{true} and  $\texttt{step} \geq n_{\rm msdstart}$, \texttt{FIRE()} is called once every $n_{\rm min}$ timesteps.
\item \texttt{swapmove()} is called once every $n_{\rm swap}$ timesteps provided $p > 0$ (Section \ref{subsec:swap}).
\item Newton's equations of motion are integrated forward in time by the increment $\Delta t = n_{\rm bet}dt$ by calling \texttt{leapfrog()} $n_{\rm bet}$ times.
\end{itemize}
Here \texttt{step} indicates the timestep \#.
$n_{\rm bet}$ is analogous to LAMMPS' ``delay'' parameter \cite{lammps}; large values reduce simulation runtimes slightly but run the risk of insufficiently-frequent VL builds that lead to lost-atom crashes.
For the $T = 0.25\tilde{\epsilon}/k_B$ runs described below, we found that $n_{\rm bet} = 5$ was sufficiently low to ensure stability of the runs.
Lower $T$ allow larger $n_{\rm bet}$ values.

Newton's equation of motion are integrated via the ``kick-drift-kick'' variant of the leapfrog method \cite{leapfrog}.
For maximum efficiency, thermostatting in NVT simulations (see Section \ref{subsec:auxfcns}) is implemented as part of \texttt{leapfrog()}'s second kick (i.e.\ its second velocity update); this improves cache locality.
Force evaluation within \texttt{getforce()} is handled in standard fashion, with one exception; for $d > 5$, the pair distance calculation 
\begin{equation}
r_{ij} = \sum_{k = 1}^d  (\vec{r}_j - \vec{r}_i)\cdot \hat{x}_i ,
\end{equation}
where $\hat{x}_i$ is the unit vector pointing along the $i$th spatial axis, is truncated before its completion if the partial sum exceeds the cutoff radius.

Efficient energy minimization is implemented using the semi-implicit Euler variant of the  ``FIRE 2.0'' update \cite{guenole20} to the original FIRE algorithm \cite{bitzek06}.
\texttt{hdMD}'s \texttt{FIRE()} routine takes two arguments: $E_{\rm thres}$ and $F_{\rm thres}$.
Minimization stops when the average pair energy drops below $E_{\rm thres}$, the fractional pair energy drop 
\begin{equation}
F = \bigg{ | } \displaystyle\frac{E_{i+1} - E_i}{E_i} \bigg{ | }
\label{eq:Fthres}
\end{equation}
is smaller than $F_{\rm thres}$ for ten consecutive iteration steps, or the iteration count reaches \texttt{maxiter}.
During the abovementioned ``pushoff'', complete energy minimization is unnecessary, so we set $E_{\rm thres} = \epsilon$.
\texttt{hdMD} also includes \texttt{steepest()}, a standard adapative-timestep gradient-descent energy minimization routine which stops when the average pair energy drops below $E_{\rm thres}$ (its only argument).

\subsection{Cell setup and Verlet-list building}
\label{subsec:NLbuilds}

As discussed above, \texttt{hdMD}'s simulation cells are hypercubes of side length $L$, divided into $\mathcal{N}_{\rm sc} = n_{\rm cell}^d$ cubic subcells with side length $L_{\rm sc} \equiv L/n_{\rm cell}$. 
Here
\begin{equation}
n_{\rm cell} = \rm{floor} \left(  \displaystyle\frac{L}{ \mathcal{R}^{\frac{1}{2d}}\tilde{\sigma} + s} \right),
\label{eq:setnc}
\end{equation}
where $\mathcal{R}^{\frac{1}{2d}}\tilde{\sigma}$ is the diameter of the largest particles, $s$ is the skin depth, and $\rm{floor}[x]$ rounds $x$ downward to the nearest integer, e.g.\ $\rm{floor}[12.3] = 12$.
All particles' VLs are rebuilt any time \texttt{needsrebuild()} finds that at least $0.1\%$ of particles have moved by at least $s/2$ or any particles have moved by at least $s$ since the previous build.
Partial VL builds wherein only the VLs of particles that have moved by at least $s/2$ and those in neighboring subcells are rebuilt \cite{yao04} increased runtimes for simulations like those reported below, so this capability was not included in the final code.
We found that $s = .25\tilde{\sigma}$ works well for a wide range of conditions and use this value in all simulations reported below.

The maximum cutoff radius for inclusion in any particle's VL, $r_c^{\rm max} = \mathcal{R}^{\frac{1}{2d}}\tilde{\sigma} + s$, is the maximum interaction range, (i.e.\ the maximum interparticle distance for which nonzero forces can arise) plus $s$.
Thus $L_{\rm sc}$ is equal to or slightly larger than $r_c^{\rm max}$. 
Many modern MD codes employ smaller subcells of side length equal to or slightly larger than $r_c^{\rm max}/2$ and link each subcell to its $5^d - 1$ nearest neighbors (as opposed to its $3^d - 1$ adjacent neighbors as discussed above).
Implementing these smaller subcells multiplies the number of pair distances which must be evaluated during VL builds by a factor $\sim (5/6)^d$ and often speeds up VL building \cite{plimpton95,welling11}.
However, it also leads to increased overhead because  $\mathcal{N}_{\rm sc}$ increases by a factor $\sim 2^d$.
We find that for the short-ranged interactions employed here, the reduction in the number of pair-distance calculations is outweighed by the increased overhead.
However, the smaller-subcell approach is likely faster for sufficiently-long-ranged interactions and can be implemented by editing two lines of code in \texttt{main()}.

The linked subcells and machinery for VL-building are set up at the beginning of any simulation.
Subcells centered at 
\begin{equation}
\vec{R} = \sum_{i = 1}^d \left( -\displaystyle\frac{1}{2} + \displaystyle\frac{C_i+1/2}{n_{\rm cell}} \right)L \hat{x}_i ,
\end{equation}
where $C_i$ is an integer satisfying $0 \leq C_i < n_{\rm cell}$, and the $-1/2$ term is present because the simulation cell is centered at the origin, are given the index
\begin{equation}
\mathcal{I} = \sum_{i = 1}^d C_i n_{\rm cell}^{d-i}.
\end{equation}
Thus each subcell has a unique $\mathcal{I} \in \{0,1,2,...,\mathcal{N}_{\rm sc}-1 \}$ as well as a unique set of $\{ C_i : i = 1,2,...,d \}$.
These are defined when \texttt{setupsupercell()} is called at the beginning of a simulation.

\begin{figure}
\includegraphics[width=2.75in]{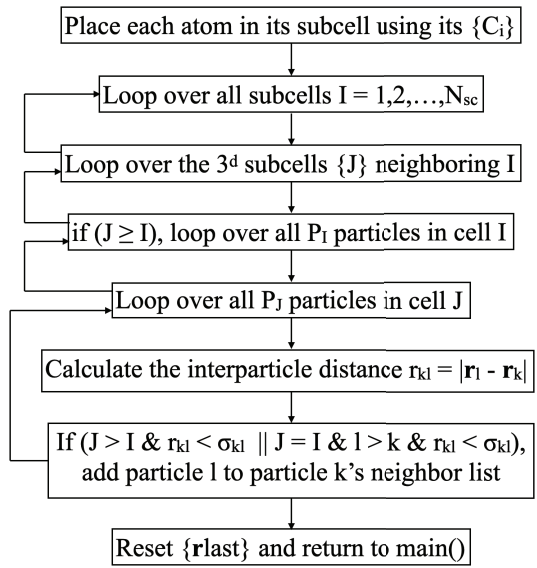}
\caption{Basic structure of the \texttt{buildneighborlist()} routine. 
Parallelization is achieved by dividing its outermost loop equally amongst the $n_{\rm threads}$ concurrent threads.}
\label{fig:VLflow}
\end{figure}

Building the VL for any given atom requires calculating the distances between it and all other atoms in its subcell as well as the $3^d - 1$ neighboring subcells.
This can be accomplished in many different ways.
After considerable trial and error, we found that the fastest algorithm is schematically described by Figure \ref{fig:VLflow}.
The basic features of this algorithm are standard for molecular simulations \cite{frenkel02}, but we found several details of this approach that are standard and work well for $d \leq 3$ perform poorly in higher $d$ and needed to be modified as detailed below.

First, implementing the outer single loop over all cells  $\mathcal{I} = 0,1,...,\mathcal{N}_{\rm sc}-1$ rather than a $d$-deep loop over the $\{ C_i \}$ both provided a substantial speedup and allowed construction of a \texttt{buildneighborlist()} routine in which $d$ is a parameter and thus works for arbitrary $d \geq 2$.
A comparable speedup was achieved by implementing a single second-from-outer loop over each subcell's $3^d$ linked subcells rather than the more geometrically intuitive approach of a $d$-deep loop over the linked subcells' $\Delta C_i = -1,0,1$ for each $i = 1,2,...,d$.

Second, while the most intuitively obvious way of handling the linked-subcell structure is to store it in static memory by identifying each subcell's $3^d - 1$ neighboring subcells at the start of the simulation and including pointers to each neighbor in each subcell's data structure (e.g.\ as a private variable in a ``subcell'' class object in C++) works well for $d \leq 3$, this approach has a substantial memory cost that significantly increases simulation runtimes in larger $d$.
We found that substantially better perfomance is achieved when each subcell's neighboring subcells are identified in the first step of the outermost loop in Fig.\ \ref{fig:VLflow}, i.e.\ each time the VL is built.

Third, we found that implementing the outer double loop over subcells and inner double loop over particles in these cells provides a substantial speedup relative to alternative loop orderings, and this speedup increases rapidly with increasing $d$.
For example, we found that the runtimes of simulations like the $d = 6$ $\phi = \phi_d$ runs discussed in Sections \ref{sec:bench}-\ref{sec:gd36} are $\sim 40\%$ larger when the order of the second and third loops in  Fig.\ \ref{fig:VLflow} is reversed.

The procedure described above requires $n_{\rm cell} \geq 3$ to function correctly, and has an $\mathcal{O}(N^2)$ computational cost when $n_{\rm cell} = 3$.
Therefore we also included a more efficient $\mathcal{O}(N^2)$ VL-building routine \texttt{buildNsq()} that replaces it whenever $\rm{floor} (  L[\mathcal{R}^{\frac{1}{2d}}\tilde{\sigma} + s]^{-1} ) \leq 3$.
\texttt{buildNsq()} eliminates the subcells entirely and performs only the bottom three steps depicted in Fig. \ref{fig:VLflow}.
For the remainder of this paper, we focus on larger systems where $n_{\rm cell} > 3$.

\subsection{SWAP Monte Carlo}
\label{subsec:swap}

Over the past decade, SWAP Monte Carlo, which speeds equilibration of deeply supercooled liquids by many orders of magnitude by exchanging particles' diameters \cite{grigera01,ninarello17}, has revolutionized simulations of the glass/jamming transition.
For example, showing that SWAP allows equilibration of hard-sphere liquids with $\phi > \phi_{\rm MRJ}$ \cite{torquato00} proved that $\phi_{\rm MRJ}$ is not the endpoint of the equilibrium-liquid branch of their phase diagram \cite{berthier16}.
The algorithm was recently extended to $d > 3$; Ref.\ \cite{berthier19c} showed that while the dynamical speedup it produces decreases exponentially with increasing $d$, it remains substantial for $d$ as large as $8$.
Therefore we included SWAP capability in \texttt{hdMD}, and describe it  here.

Once every $n_{\rm swap}$ timesteps, the \texttt{main()} routine calls \texttt{swapmove()}.
This function controls all aspects of SWAPping.
First it generates lists of the atom indices $\{ i \}$ and $\{ j \}$ for which the swaps $\sigma_i \leftrightarrow \sigma_j$, $\epsilon_i \leftrightarrow \epsilon_j$, and $m_i  \leftrightarrow m_j$ will be attempted.
Both $\{ i \}$ and $\{ j \}$ are of length $pN$ (Sec.\ \ref{subsec:integration}); the included indices are chosen randomly.
Next it generates the VLs for these sets using the subroutine \texttt{buildNLs()}.
This routine operates differently than \texttt{buildneighborlist()} for two reasons.
First, to properly calculate potential energy changes, \textit{full} VLs (as opposed to the half-VLs outlined in Section \ref{subsec:NLbuilds}) are required.
Second, there is no reason to loop over all  $\mathcal{N}_{\rm sc}$ subcells during this process.  
Instead, the subcell indices $\{ \mathcal{I}_i \}$ and $\{ \mathcal{I}_j \}$ are identified in advance.
This process, as well as the actual population of the VLs $\{ \mathcal{L}_i \}$ and $\{ \mathcal{L}_j \}$, are parallelized.
Specifically, the VL-building loops of length $pN$ over the $\{ i \}$ and $\{ j \}$ are divided equally amongst the $n_{\rm threads}$ threads.

After the VLs are built, the actual swapping is executed serially, by the master thread.
The total potential energy associated with interactions involving particles $i$ and $j$ is initially
\begin{equation}
E_b = \displaystyle\sum_k U_a(\epsilon_{ik}, \sigma_{ik},r_{ik}) +   \displaystyle\sum_l U_a(\epsilon_{jl}, \sigma_{jl},r_{jl}),
\label{eq:ebefore}
\end{equation}
where the sums are respectively over the particles $\{ k \}$ and $\{ l \}$ neighboring them.
This energy becomes 
\begin{equation}
E_a = \displaystyle\sum_k U_a(\epsilon_{jk}, \sigma_{jk},r_{ik}) +   \displaystyle\sum_l U_a(\epsilon_{il}, \sigma_{il},r_{jl}).
\label{eq:ebefore}
\end{equation}
if the swap is accepted, and the particles' velocities are rescaled to conserve kinetic energy.
Swaps are accepted or rejected using the standard Metropolis criteria.

\subsection{Temperature and pressure control, thermodynamics metrics}
\label{subsec:auxfcns}

While the results presented below are all from NVT simulations, \texttt{hdMD} also includes a Berendsen barostat \cite{berendsen84} that allows NPT simulations to be performed.
The target pressure $P_{\rm targ}$ and  target temperature $T_{\rm targ}$ can be set at the initiation of the MD run, or can be ramped by resetting them within \texttt{main()}'s main loop.
Every time \texttt{Berendsen()} is called, it checks whether  $n_{\rm cell}$ (Eq.\ \ref{eq:setnc}) has changed since the last time it was called, and
if it has, calls \texttt{setupsupercell()} and then \texttt{buildneighborlist()} to repopulate the subcells.
As mentioned above, for NVT simulations (i.e.\ when \texttt{baroflag} is set to \texttt{false}), Berendsen thermostatting is implemented within \texttt{leapfrog()}.

\texttt{hdMD}'s \texttt{getthermo()} routine calculates standard thermodynamic quantities like the temperature, pair energy $E_{\rm pair}$ from Eq.\ \ref{eq:morsepot} or its user-defined replacement, pressure $P$, and average coordination number $Z$.
Calculation and output of these quantities is performed once every $n_{\rm thermo}$ steps.
Since \texttt{hdMD} will likely be used primarily for glass/jamming-transition-related studies, it also (by default) calculates particles' mean squared displacements $\langle (\Delta \vec{r})^2 \rangle$ and mean quartic displacements $\langle (\Delta \vec{r})^4 \rangle$ since step $n_{\rm msdstart}$, the non-Gaussian parameter 
\begin{equation}
\alpha_2 = \displaystyle\frac{d \langle (\Delta \vec{r})^4 \rangle}{(d+2) \langle (\Delta \vec{r})^2 \rangle^2 } - 1
\label{eq:ngp}
\end{equation}
and overlap function
\begin{equation}
f_{\rm ov} = N^{-1} \displaystyle\sum_{i = 1}^N \Theta\left( 0.1d\tilde{\sigma} - |\Delta \vec{r}_i | \right)
\label{eq:fov}
\end{equation}
Eq.\ \ref{eq:ngp} is the $d$-dimensional generalization \cite{charbonneau12} of the usual 3D expression for $\alpha_2$ ($3\langle (\Delta \vec{r})^4 \rangle/5  \langle (\Delta \vec{r})^2 \rangle^2 - 1$).
In Eq.\ \ref{eq:fov}, $\Theta$ is the Heaviside step function, and $f_{\rm ov}$ is a simple, commonly used metric for particle relaxation in deeply supercooled liquids.
It varies continuously from $1$ to zero as particles hop away from their initial positions, and provides roughly the same information as the self-intermediate scattering function $F_s(q,t)$ evaluated at $q = 2\pi/\tilde{\sigma}$ \cite{ninarello17}.  
Finally, \texttt{hdMD}'s  \texttt{getvanHove()} routine calculates the self part of the van Hove correlation function
\begin{equation}
G_s(r,t) =  \displaystyle\frac{1}{N} \displaystyle\sum_{i = 1}^N \delta\left( |\vec{r}_i(t) - \vec{r}_i(0) | - r \right),
\label{eq:VGC}
\end{equation}
where $\delta$(x) is the Dirac delta function.
Note that \texttt{hdMD}'s main loop is structured in such a way that additional periodic calculations of other thermodynamic quantities can easily be added by the user.

\section{Performance of simulations at $\phi \simeq \phi_d$ for $3 \leq d \leq 6$}
\label{sec:bench}

\subsection{Theoretical Background}
\label{subsec:theoback}

The dynamical glass transition packing fraction $\phi_d$ is defined as the packing fraction at which a supercooled $d$-dimensional hard-sphere liquid's diffusivity would drop to zero \textit{if there were no hopping motion} \cite{kirkpatrick89}.
Hopping makes the actual packing fraction at which diffusivity drops to zero substantially higher, but examining systems with $\phi \simeq \phi_d$ has proven very fruitful  \cite{charbonneau12,charbonneau14c} for understanding the ways in which finite-$d$ supercooled liquids and glasses differ from their exactly-solvable, mean-field counterparts \cite{parisi10,charbonneau14}.
In this section, we examine \texttt{hdMD}'s scalability and parallel efficiency for simulations of systems at packing fractions and temperatures that map to $\phi_d$ but have orders-of-magnitude-larger $N$ than have been employed in previous $d > 3$ studies \cite{lue06,skoge06,vanMeel09,vanMeel09b,charbonneau11,charbonneau12,charbonneau13b,charbonneau14c,charbonneau14,adhikari21,berthier19c,berthier20,morse21,charbonneau21}.  
We show that its performance (as judged by these metrics) is comparable to those achieved by popular $d = 3$ MD codes \cite{lammps,gromacs,namd,amber,hoomd,rumd}.
Then we examine how \texttt{hdMD}'s performance varies with $d$.
We show that in the large-$N$ limit it is nearly optimal, i.e.\ the runtime per pair distance calculation is nearly $d$-independent.

Results from simulations employing the Morse potential can be compared to hard-sphere results by considering systems at the same effective packing fraction $\phi_{\rm eff} = \phi \sigma_{\rm eff}^d$, where the temperature-dependent effective hard-sphere diameter \cite{weeks71} is
\begin{equation}
\sigma_{\rm eff}(a,T) = \int_{0}^{\tilde{\sigma}} \left[1-\exp{(-U_a(r)/k_B T)}\right] dr.
\label{eq:seff}
\end{equation}
More accurate mapping to hard sphere results can be achieved using a refined version of this method \cite{schmiedeberg11}, but since the primary focus of this paper is demonstrating the utility of \texttt{hdMD} rather than precisely matching hard-sphere results, we use Eq.\ \ref{eq:seff} to estimate the dynamical-glass-transition packing fractions $\phi_d^*$ for $a = 30$ and $T = 0.25\epsilon/k_B$ (Table \ref{tab:phiefftab}).

\begin{table}[htbp]
\caption{Dynamical-glass-transition packing fractions $\phi_d$ for hard-spheres \cite{charbonneau12}, the associated $\sigma_{\rm eff}^d$ values for $a = 30$ and $T = 0.25\epsilon/k_B$ (Eq.\ \ref{eq:seff}), and the packing fractions $\phi^*_d = [\sigma_{eff}(30,0.25\epsilon/k_B)]^{-d} \phi_d$ employed in the simulations discussed in this section.}  
\begin{ruledtabular}
\begin{tabular}{lccc} 
$d$ & $\phi_d$ & $[\sigma_{\rm eff}(30,0.25\epsilon/k_B)]^d$ & $\phi^*_d$ \\
3 & 0.5770 & 0.9650 & 0.5980 \\
4 & 0.4036 & 0.9536 & 0.4232 \\
5 & 0.2683 & 0.9423 & 0.2847 \\
6 & 0.1723 & 0.9312 & 0.1850
\end{tabular}
\end{ruledtabular}
\label{tab:phiefftab}
\end{table}

\subsection{Scaling of runtimes with $N$, $n_{\rm threads}$, and $d$}
\label{subsec:scal}

\begin{figure}[h!]
\includegraphics[width=3in]{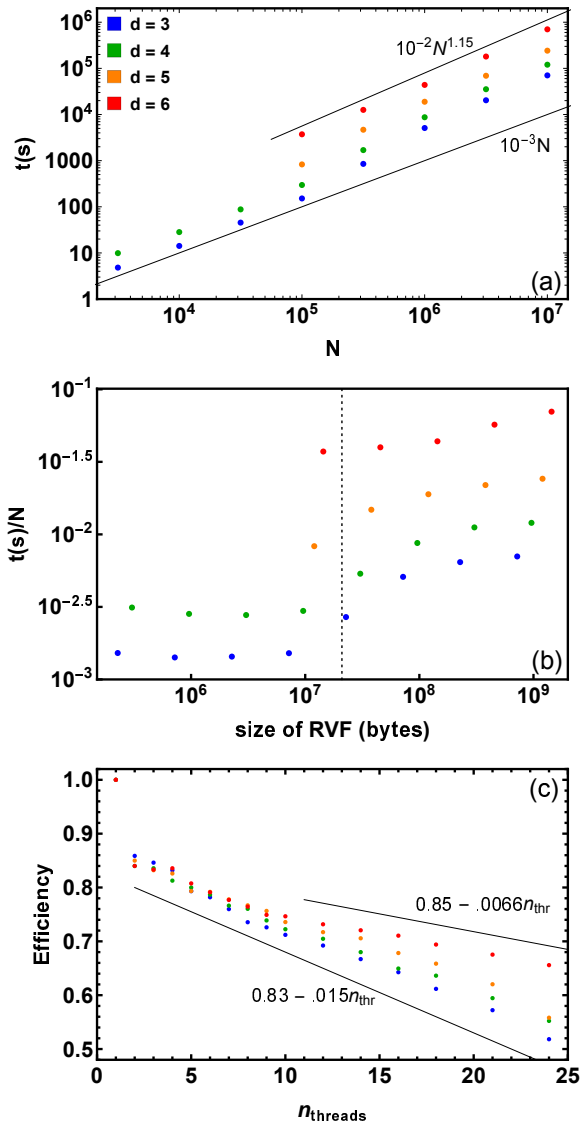}
\caption{ Scaling of performance with $N$ and $n_{\rm threads}$ for $3 \leq d \leq 6$.  Runtimes for panels (a-b) are for $n_{\rm threads} = 8$ simulations run on an iMac with a single 10-core Intel Core i9 CPU (3.6GHz).  
Results for $N < 10^5$ and $d > 4$ are not shown here because these $N$ give $n_{\rm cell} \leq 4$ and the scaling of the time devoted to VL builds is closer to $\mathcal{O}(N^2)$ than $\mathcal{O}(N)$.
Runtimes for panel (c) are for $N = 10^6$ simulations on cluster nodes with two 12-core Intel Xeon E5-2650 CPUs (2.2GHz).  The larger efficiencies for $n_{\rm threads} > 12$ probably arise from distributing the computational effort to more than CPU socket.
Solid lines in  panels (a, c) are guides to the eye, and the dotted vertical line in panel (b) indicates the iMac's L3 cache size (20MB).}
\label{fig:scalfig1}
\end{figure}

Figure \ref{fig:scalfig1} summarizes \texttt{hdMD}'s scalability and parallel efficiency for short ($100\tau$) simulations of supercooled liquids at this temperature and $\phi = \phi_d^*$ for $3 \leq d \leq 6$.
All simulations include 10 FIRE energy minimizations and 100 SWAP-MC passes, respectively performed  once every $10\tau$ and once every $\tau$ after the simulation begins.
The FIRE minimizations employ $E_{\rm thres} = 10^{-25}\tilde{\epsilon}$, and the SWAP passes employ $p = 0.1$, which is close to the optimal value \cite{ninarello17}.
For $a = 30$, $dt = 0.0016\tau$, so there are 625 MD timesteps per $\tau$ and thus 62500 total timesteps in each simulation.

Panels (a-b) show  how runtimes increase with $N$ for $10^{3.5} \leq N \leq 10^7$ for $3 \leq d \leq 4$ and  $10^5 \leq N \leq 10^7$ for $5 \leq d \leq 6$.
For $N \leq 10^5$, runtimes scale as $N^y$ with $y \simeq 1.04$, which is very close to the optimal linear scaling.
As $dN$ increases beyond $\sim 10^6$, the particles' \texttt{rvf} array (Sec.\ \ref{subsec:integration}) can no longer easily fit within the CPU's L3 cache, and calls to \texttt{getforce()} produce more and more cache misses.
This worsens the runtime scaling to $1.1 \lesssim y \lesssim 1.15$, which is suboptimal, but only slightly so.
Some standard ($d = 3$) MD codes that are optimized for large $N$ (e.g.\ LAMMPS \cite{lammps}) periodically re-order particle ids by the particles' positions along one dimension \cite{yao04,meloni07}.
This improves their large-$N$ scaling by improving cache locality.
Since $N = 10^5$ should be large enough for most glass-transition-related studies likely to be conducted in the next few years, we have not yet implemented this feature in  \texttt{hdMD}, but it can easily be added; we may do so in the near future.

Panel (c) shows how parallel efficiency (PE) decreases as $n_{\rm threads}$ increases for $N = 10^6$ simulations on a typical mid-2010s dual-socket cluster node.
For $n_{\rm threads} = 2$, PE values are $\sim 85\%$.
This is typical for numerical applications of OpenMP; PE is never $100\%$ because there is a $\sim 10\mu s$ overhead associated with parallelizing any \texttt{for} loop.
As discussed above, some of the tasks executed during these simulations, e.g.\ the assignment of particles to subcells and the SWAP moves, are not parallelized or readily parallelizable.
Fortunately, PE drops only slowly as $n_{\rm threads}$ increases.
As shown in the figure, a very loose lower bound for PE is $(83 - 1.5n_{\rm threads})\%$, and higher-$d$ simulations have considerably larger PE.
For large $N$, \texttt{hdMD}'s  efficiency-limiting factor appears to be the memory-bound force-evaluations; increasing $n_{\rm threads}$ increases the rate of cache misses in \texttt{getforce()}.
Thus $\rm{PE}(n_{\rm threads})$ might also be improved by implementing particle-id reordering.
We leave this for future work, but emphasize that since VL-building is less memory-bound than integration of the EOM, $\rm{PE}(n_{\rm threads})$ actually increases with increasing $d$.

\begin{figure}
\includegraphics[width=2.9in]{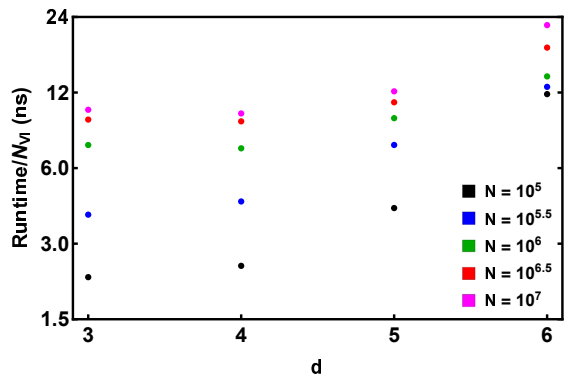}
\caption{Scaling of performance with $d$. Runtimes are for $n_{\rm threads} = 8$ simulations run on an iMac with a single 10-core Intel Core i9 CPU (3.6GHz).}
\label{fig:scalfig2}
\end{figure}

Next we discuss how runtimes for fixed $N$ and $n_{\rm threads}$ vary with $d$.  
In general, runtimes for fixed $\hat{\phi} = 2^d\phi$ must increase with $d$ for two reasons.
First, the effort required for VL building scales with $\mathcal{N}_{\rm Vl} \sim3^d$ as discussed above.
Second, the size of particles VLs in these runs scales as $\rho_d v(d,\sigma + s)$ (Eq.\ \ref{eq:vofsigma}).
For these $\phi = \phi^*_d$ runs, $\mathcal{N}_{\rm Vl} = 11.0,\ 19.4,\ 31.9$, and $50.4$ in $d = 3-6$.
Thus the runtime \textit{per pair distance calculation} in \texttt{getforce()}, i.e.\ $t/[n_{\rm steps} \mathcal{N}_{\rm Vl}]$, is a good metric for comparing runtimes across different $d$.
Results for all $N \geq 10^5$ systems are shown in Figure \ref{fig:scalfig2}.
For the smaller $N$, $t/\mathcal{N}_{\rm Vl}$ increases rapidly with $d$ because $n_{\rm cell}$ drops to low values (e.g.\  $n_{\rm cell} = 4$ for $N = 10^5$ and $d = 6$) and thus a large fraction of the $N$ particles must be searched over during the building of each particle's VL.
For $N \geq 10^6$, however, $t/\mathcal{N}_{\rm Vl}$ increases far slower, only increasing by a factor of $\sim 2$ as $d$ increases from $3$ to $6$.
This indicates that for large simulations the scaling of \texttt{hdMD} runtimes with $d$ is nearly optimal given that pair-distance calculations are the rate-limiting factor.

\begin{table}[h]
\begin{ruledtabular}
\caption{Runtime percentages for $n_{\rm threads} = 8$  simulations with  $N = 10^5$ (top rows) and $N = 10^7$ (bottom rows).}
\begin{tabular}{cccccc}
$d$ & Leapfrog & VL-building & Minimizations & SWAP & Other \\
\hline
3 & 83.9 & 11.1 & 2.1 & 1.2 & 1.7 \\
4 & 72.0 & 22.4 & 1.5 & 2.9 & 1.2 \\
5 & 47.5 & 45.9 & 1.0 & 4.6 & 1.0 \\
6 & 26.0 & 67.2 & 0.7 & 5.3 & 0.7 \\
\hline
3 & 90.6 & 5.5 & 1.5 & 0.6 & 1.7 \\
4 & 88.2 & 7.6 & 1.5 & 1.4 & 1.4 \\
5 & 72.2 & 20.1 & 1.3 & 3.5 & 3.0 \\
6 & 42.8 & 43.5 & 0.8 & 4.2 & 1.8 
\end{tabular}
\end{ruledtabular}
\label{tab:timebyroutine}
\end{table}

The trends shown in Fig.\ \ref{fig:scalfig2} can be understood by examining in greater detail how simulation runtimes are divided among \texttt{hdMD}'s various routines.
Table IV lists the runtime percentages devoted to leapfrog integration, VL building, energy minimizations, and SWAP for $N = 10^5$ and $N = 10^7$, which are representative of simulations of moderately-sized and very-large systems.
For both $N$, the percentage of simulation runtimes spent in VL-building (leapfrog integration) increases (decreases) rapidly with increasing $d$.
However, the fractional increases/decreases in these percentages as $d$ increases from $3$ to $6$ are far greater for $N = 10^5$ than for $N = 10^7$.
This explains why the runtimes per pair distance calculation increase faster with $d$ when $N$ is smaller and vice versa.

Taken together, the above results show that \texttt{hdMD} makes large-$N$ simulations practical in $d$ up to 6, even with modest computational resources.
In the following section, we demonstrate how this feature can be exploited to obtain novel physics results.

\section{Heterogeneous dynamics in deeply supercooled $d = 6$ liquids}
\label{sec:gd36}

\subsection{Breakdown of the Stokes-Einstein relation}
\label{subsec:SER}

Three recent simulation studies \cite{charbonneau12,charbonneau13b,adhikari21} of dynamical heterogeneity in supercooled liquids in $3 \leq d \leq 10$ have reported that  it weakens with increasing $d$.
However, as mentioned in the Introduction, it may be that the heterogeneous dynamics in these simulations \textit{artificially} weakened with increasing $d$ because they employed small fixed $N < 10^4$ and periodic boundary conditions with $L \sim N^{1/d}$ that may have dropped below the characteristic size $\mathcal{L}$ of the liquids' spatial heterogeneities \cite{berthier12}.
Whether this is so can be determined by simulating liquids with fixed $L$ rather than fixed $N$ over a comparable range of $d$, taking care that $L > \mathcal{L}$ for all $d$ \cite{eaves09}.
As described above, \texttt{hdMD}'s efficient parallel implementation makes it well suited to doing so.
Here we motivate such studies by showing that $d = 6$ supercooled liquids can be substantially more heterogeneous that previously reported, and provide some initial evidence that the strengths of the Stokes-Einstein-relation (SER) breakdowns reported in  Refs.\ \cite{charbonneau12,charbonneau13b,adhikari21} were artificially suppressed by the small system sizes they employed.

\begin{figure}
\includegraphics[width=3in]{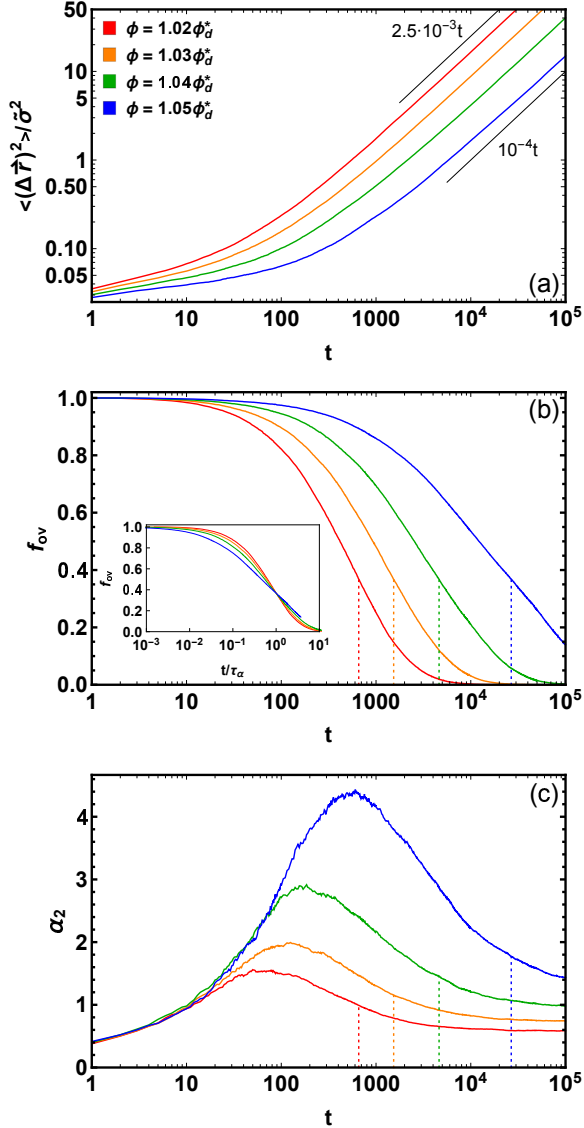}
\caption{ Dynamics of supercooled $d = 6$ liquids for $1.02\phi_d^* \leq \phi \leq 1.05\phi_d^*$ and $k_B T = 0.25\tilde{\epsilon}$.  Panels (a-c) respectively show the mean-squared displacement, overlap parameter $f_{\rm ov}$ (Eq.\ \ref{eq:fov}), and non-Gaussian parameter (Eq.\ \ref{eq:ngp}).  Straight gray lines in panel (a) are guides to the eye, and dotted vertical lines in panels (b-c) indicate $t = \tau_\alpha$.  All times are given in units of  $\tau = \sqrt{\tilde{m}\tilde{\sigma}^2/\tilde{\epsilon}}$ (Sec.\ \ref{subsec:model}). }
\label{fig:glasdyn}
\end{figure}

Figure \ref{fig:glasdyn} shows dynamical results for supercooled $d = 6$ liquids with $1.02\phi_d^* \leq \phi \leq 1.05\phi_d^*$ (Section \ref{subsec:theoback}, Table \ref{tab:phiefftab}).
To obtain a cleaner comparison to the results of Refs.\ \cite{charbonneau12,charbonneau13b,adhikari21}, all particles in these simulations were given equal masses ($m_i = \tilde{m} = 1$).
All systems have  $N = 10^5$ and were SWAP-equilibrated at  $k_B T = 0.25\tilde{\epsilon}$ for at least $1.75{\times} 10^4 \tau$.
Following the equilibration runs, SWAP was turned off and systems were evolved forward in time for another $\mathcal{T} = 10^5\tau$ (also at  $k_B T = 0.25\tilde{\epsilon}$).
No ensemble or time averaging was performed.

Panel (a) shows particles' mean squared displacements $\langle [\Delta \vec{r}(t) ]^2 \rangle$.
The increasing MSD plateau length with increasing $\phi$ is typical for supercooled liquids \cite{kob95}, as is the gradual increase in the slope  $\zeta(t) \equiv d \ln[\langle \Delta \vec{r}(t) ]^2 \rangle]/d\ln(t)$ towards $1$ as systems approach the Brownian-diffusive regime.
All systems reach this regime well before $t = \mathcal{T}$, and the values of  $\langle [\Delta \vec{r}(\mathcal{T}) ]^2 \rangle/\tilde{\sigma}^2$ are all well above $1$, indicating that typical particles have hopped multiple times by the end of these simulations even for $\phi = 1.05\phi_d^*$.

Panel (b) shows particles' overlap parameter $f_{\rm ov}(t)$.
Again, all results are typical for supercooled liquids.
Following Refs.\ \cite{ninarello17,adhikari21}, we define the alpha relaxation times ($\tau_\alpha$) for these liquids using the criterion $f_{\rm ov}(\tau_\alpha) = e^{-1}$.
Numerical results for $\tau_{\alpha}$, the diffusion coefficients $D = \lim_{t \to \infty} \langle [\Delta \vec{r}(t) ]^2 \rangle/2dt$,  and several related quantities are shown for a wider range of $\phi$ in Table V.
As shown in the inset, the $f_{\rm ov}(t)$ curves do not collapse when plotted vs.\ $\tau/\tau_\alpha$, indicating that time-density superposition breaks down in these systems.
$f_{\rm ov}(t/\tau)$ decreases faster as $\phi$ increases because the dynamics of small and large particles have decoupled; larger particles' mobility is decreasing faster with increaing $\phi$ than that of their smaller counterparts.
Such decoupling has long been associated with dynamical heterogeneity \cite{kob95,ding06,saltzman06}.

Panel (c) shows results for the non-Gaussian parameter $\alpha_2(t)$.
As expected from previous $d = 3$ studies \cite{kob95}, results for different $\phi$ fall on a common curve at small $t$ and exhibit maxima $\alpha_{2, \rm{max}}(\phi)$ at times $\tau^*(\phi)$ that increase with $\phi$ slower than $\tau_{\alpha}(\phi)$.
As expected for relatively-low-temperature liquids, $\alpha_{2,{\rm max}} \sim  \tau_{\alpha}^x$ with $x \simeq 0.3$ \cite{wang18,footnandi,nandi21}.
Intriguingly, the $\alpha_2(\tau_\alpha)$ values increase roughly logarithmically with $\tau_\alpha$ as $\phi$ increases rather than as a power law, i.e.\ $\alpha_2(\tau_\alpha) \sim \ln(\tau_\alpha)$.

At longer times, rather than trending back to zero as is the case when monodisperse systems are considered or $\alpha_2$ is calculated using only one component of a bidisperse mixture \cite{kob95,charbonneau12,charbonneau13b,adhikari21}, all systems' $\alpha_2(\phi,t)$ decay very slowly (over timescales of order $10^2 \tau_{\alpha}$) towards finite plateau values $\alpha_{2,\infty}$ that increase rapidly with $\phi$.
Finite $\alpha_{2,\infty}$ are expected since smaller particles have larger diffusion coefficients $D_i = \tilde{D}(\sigma_i)$, where $\tilde{D}$ is an \textit{a priori} unknown function that captures the particle-size dependence of diffusivity.
Quantitatively, one expects  \cite{abete08}
\begin{equation}
\alpha_{2,\infty} = \displaystyle\frac{\langle \tilde{D}^2 \rangle}{ \langle \tilde{D} \rangle^2 } - 1 ,
\end{equation}
where
\begin{equation}
\langle \tilde{D}^n \rangle = \displaystyle\int_{\sigma_{\rm min}}^{\sigma_{\rm max}} P(\sigma) [\tilde{D}(\sigma)]^n d\sigma .
\end{equation}
For the particle-size distribution employed here [i.e.\ $P(\sigma) \equiv P(6,6,\sigma)$ from Eq.\ \ref{eq:sigmadist}], assuming all particles obey the classical relation $\tilde{D}(\sigma) \propto k_BT(\eta \sigma)^{-1}$ [where $\eta$ is viscosity] predicts $\alpha_{2,\infty} \simeq .047$.
Actual $\alpha_{2,\infty}$ values are much larger and increase rapidly with increasing $\phi$, consistent with the well-known result that deviations of $\tilde{D}(\sigma)$ from this formula strengthen with increasing $\phi$ or decreasing $T$ \cite{rossler90}.
This contributes to the abovementioned breakdown of $t$-$\phi$ superposition.
One expects, based on previous studies of polydisperse $d = 3$ supercooled liquids performed as far back as the mid-2000s \cite{murarka03,kumar06}, that it will  also contribute to the SER breakdown, indicated by  the increase in $D  \tau_\alpha$ with increasing $\phi$, that occurs as particle motion becomes increasingly hopping-dominated.

\begin{table}[htbp]
\begin{ruledtabular}
\caption{Measures of mobility and dynamical heterogeneity in supercooled $d = 6$ liquids. 
$\tau_\alpha$ and $\tau^*$ are given in units of  $\tau = \sqrt{\tilde{m}\tilde{\sigma}^2/\tilde{\epsilon}}$ (Sec.\ \ref{subsec:model}).   
Results for $\phi \leq 1.04\phi_d^*$ are well fit by  $D \sim (\tilde{\phi}_d - \phi)^{\gamma_D}$ and $\tau_\alpha \sim (\tilde{\phi}_d - \phi)^{-\gamma_\tau}$, with $\tilde{\phi}_d  = 1.068\phi_d^* = .1976$, $\gamma_d \simeq 8/3$, and $\gamma_\tau \simeq 10/3$.
$\alpha_{2,\infty}$ values are given only for systems in which $\alpha_2(t)$ has clearly reached its plateau value by $t = 10^5\tau$.}
\begin{tabular}{ccccccc}
$\phi/\phi_d^*$ & $D \tilde{\sigma}^{-2}\tau$ & $\tau_\alpha$ & $\alpha_{2,\rm{max}}$ & $\tau^*$  & $\alpha_2(\tau_\alpha)$ & $\alpha_{2,\infty}$ \\
1.00 &  $4.11{\times} 10^{-3}$ & $2.09{\times}10^2$ & 1.06 & 30 & 0.735 & 0.41 \\
1.005 & $3.40{\times} 10^{-3}$ & $2.63{\times}10^2$ &  1.15 & 32 & 0.769  & 0.45 \\ 
1.01 & $2.67{\times} 10^{-3}$ & $3.58{\times}10^2$ & 1.24 & 59  & 0.847 & 0.49 \\
1.015 & $2.13{\times} 10^{-3}$ & $4.77{\times}10^2$ & 1.40 & 66 & 0.898 & 0.54 \\
1.02 & $1.62{\times} 10^{-3}$ & $6.57{\times}10^2$ & 1.54 & 75  & 0.992 & 0.59 \\
1.025 & $1.22{\times} 10^{-3}$ & $9.92{\times}10^2$ & 1.79 & 91 & 1.079 & -- \\
1.03 & $8.72{\times} 10^{-4}$ & $1.54{\times}10^3$ & 2.00 & 127  & 1.154 & -- \\
1.035 & $6.09{\times}10^{-4}$ & $2.58{\times}10^3$ & 2.40 & 126 & 1.321 & -- \\
1.04 & $4.02{\times} 10^{-4}$ & $4.63{\times}10^3$ & 2.93 & 158  & 1.455 & -- \\
1.045 & $2.52{\times} 10^{-4}$ &  $1.03{\times}10^4$ & 3.56 & 295 & 1.613 & -- \\
1.05 & $1.46{\times} 10^{-4}$ & $2.67{\times}10^4$ & 4.44 & 598 & 1.770 & -- \\
\end{tabular}
\end{ruledtabular}
\label{tab:d6mdh}
\end{table}

While the qualitative behaviors summarized in Figure \ref{fig:glasdyn}(c) and Table V are unremarkable in and of themselves, they \textit{are} noteworthy because they show in two distinct ways that high-$d$ supercooled liquids can be more heterogeneous than previously reported.
First, the $\alpha_{2,\rm{max}}$ values are substantially higher than any reported in Refs.\ \cite{charbonneau12,adhikari21}, neither of which showed any $\alpha_{2,\rm{max}} > 1.6$ for $d = 6$ liquids at any $T$ or $\phi$.
The recently demonstrated one-to-one correspondence between $\alpha_{2,\rm{max}}$ and the kinetic fragility $m^*$ \cite{wang18} implies that these liquids are also more fragile than any $d = 6$ liquids studied in Refs.\ \cite{charbonneau12,adhikari21}.
Second, they show that the SER violations in these liquids (as quantified via the relation $D\tau_\alpha \sim \tau_\alpha^\omega$) can be much stronger than observed in Refs.\ \cite{charbonneau13b,adhikari21}.

The quantity $\omega$ is of particular interest for its ability to shed light on the $d$-dependence of dynamical heterogeneity in supercooled liquids \cite{eaves09}.
$D$ is dominated by the fastest (smallest) particles, while $\tau_\alpha$ is primarily set by the slowest (largest) particles \cite{kumar06}.
Since $\tau_\alpha$ increases with $\phi$ faster than $D$ decreases, the product $D\tau_\alpha$ increases with both $\phi$ and $\tau_\alpha$, implying $\omega > 0$.
The strength of this effect should decrease with increasing $d$ because particles' cages become more mean-field-like \cite{charbonneau12}.
Mean-field theories predict $D \sim (\phi_d - \phi)^\gamma$ and $\tau_\alpha \sim (\phi_d - \phi)^{-\gamma}$ as $\phi$ approaches $\phi_d$ from below, implying $\omega = 0$.
Additional theoretical analyses predict that $\omega$ should vanish above the upper critical dimension $d_u = 8$ \cite{biroli07}.
Numerical results in Refs.\  \cite{charbonneau13b,adhikari21} were consistent with this hypothesis, and suggested $\omega \sim (d_u - d)$. 
On the other hand, studies of the mean-field Mari-Kurchan model \cite{charbonneau14c} showed $\omega \simeq .22$ for all $2 \leq d \leq 6$, while a recent study of the kinetically constrained East model showed \cite{kim17} that $\omega$ remains finite for all $d \leq 10$ and may remain finite in the $d \to \infty$ limit, suggesting that this issue has not yet been resolved.

\begin{figure}
\includegraphics[width=2.9in]{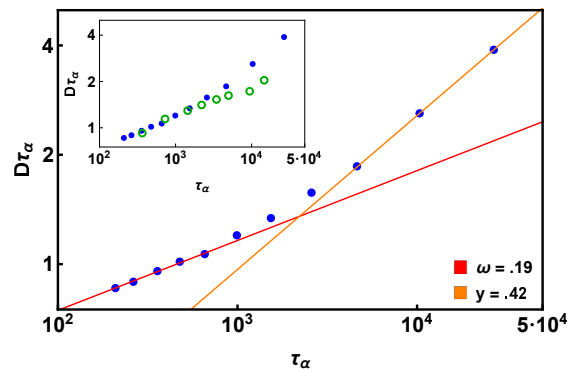}
\caption{ Breakdown of the Stokes-Einstein relation in supercooled $d = 6$ liquids at $k_B T = 0.25\tilde{\epsilon}$.  Symbols show MD data while lines show fits to $D\tau_\alpha \sim \tau_\alpha^\omega$ and $D\tau_\alpha \sim \tau_\alpha^y$.    The inset contrasts results for $N = 10^5$ (same symbols shown in the main panel) to $N = 5000$ results for selected $1.01\phi_d^* \leq \phi \leq 1.055\phi_d^*$  ({open} circles); all results for $N = 5000$ are averaged over ten independently prepared systems.
Note that the range of $\tau_\alpha$ depicted here is almost identical to that considered in Ref.\ \cite{adhikari21}.  }
\label{fig:dtau}
\end{figure}

The $D(\phi)$ and $\tau_{\alpha}(\phi)$ data shown in Table V are qualitatively consistent with those reported in many previous studies.
For $\phi$ not too close to the packing fraction $\tilde{\phi}_d$ where diffusive motion ceases,  $D \sim (\tilde{\phi}_d - \phi)^{\gamma_D}$ and $\tau \sim (\tilde{\phi}_d - \phi)^{-\gamma_\tau}$, where  $\tilde{\phi}_d$ is several percent above $\phi_d^*$ \cite{charbonneau14c} and (in contrast to mean field theories) $\gamma_\tau > \gamma_D$. 
Thus $D\tau_\alpha \sim  (\tilde{\phi}_d - \phi)^{\gamma_D  - \gamma_\tau} \sim \tau_\alpha^\omega$ with  $\omega = 1-  \gamma_D/\gamma_\tau$.
Figure \ref{fig:dtau} shows $D \tau_\alpha$ vs.\ $\tau_{\alpha}$ for  these systems.
Lower-density ($\phi \lesssim 1.02\phi_d^*$) liquids' results fall on a common curve $D \tau_\alpha \sim \tau_{\alpha}^\omega$ with $\omega \simeq 0.2$.
At higher densities, a crossover to a stronger dependence $D \tau_\alpha \sim \tau_{\alpha}^y$ is observed as systems become sluggish, consistent with Fig.\ 7(b) of Ref.\ \cite{charbonneau13b}.
Overall, the trends are the same as found in Refs.\  \cite{charbonneau13b,adhikari21}, but the $\omega$ value is more than twice as large as in these studies, which respectively found $\omega \simeq .09$  \cite{charbonneau13b} and $\omega \simeq .083$ \cite{adhikari21} in comparable $d = 6$ liquids.

There are multiple potential reasons for this difference. 
For example, here we have employed a moderate-stiffness ($a = 30$) Morse pair potential and moderate temperature.
In contrast, Refs.\ \cite{charbonneau12,charbonneau13b} employed hard spheres while Ref.\ \cite{adhikari21} employed soft harmonic spheres at very low $T$.
Thus our liquids experience thermal activation over energy barriers (absent from \cite{charbonneau12,charbonneau13b}) and substantially higher mobilities than those of \cite{adhikari21}.

Another potential reason is that our systems are much larger, with $N = 10^5$ rather than $N = 5000-8000$ as was the case in Refs.\ \cite{charbonneau12,charbonneau13b,adhikari21}.
As discussed above, one expects dynamical heterogeneity to increase with system size.
To investigate this possibility, we characterized the dynamics of $N = 5000$ systems at the same $\{ \phi \}$ and $T$.
We found that $D$ decreases with $\phi$ slightly slower in these liquids than in their $N = 10^5$ counterparts, but $\tau_\alpha$ increases substantially slower, particularly for $\phi \gtrsim 1.03\phi_d^*$.
Results for $D\tau_\alpha$ for these liquids are shown in the inset to Fig.\ \ref{fig:dtau}.
Plainly $D\tau_\alpha$ grows slower with increasing $\phi$ than in the $N = 10^5$ liquids, and as a consequence, the \textit{apparent} breakdown of the SER is weaker.
This difference presumably arises because periodic boundary conditions cap the characteristic size of cooperatively rearranging regions within a model supercooled liquid at $L$; the $N = 5000$ liquids' smaller $L$ reduces the characteristic size of their cooperatively rearranging regions and hence their $\tau_\alpha$ \cite{starr13}.
If true, this would explain these liquids' delayed crossover to the stronger $D\tau_\alpha \sim \tau_\alpha^y$ scaling.

It is reasonable to suppose that when comparing systems with fixed $N$ and $\phi/\phi_d^*$ across multiple $d$ as was done in Refs.\ \cite{charbonneau12,charbonneau13b,adhikari21}, the 
decrease in $L$ with increasing $d$ produces a comparable (artificial) reduction in the measured $D\tau_\alpha$ and perhaps also in the inferred $\omega$  \cite{footgarrahan}.
As mentioned above, this hypothesis could be tested using simulations where $L$ rather than $N$ is fixed \cite{eaves09}.
Our present focus is not to resolve this issue -- finite-size effects on the dynamics of supercooled liquids are decidedly nontrivial \cite{berthier12} -- but rather to demonstrate that \texttt{hdMD} is well-suited to doing so.

\subsection{Comparison to bidisperse systems}

A third potential reason for the larger $\alpha_{2,\rm max}$ and $\omega$ reported above is our use of continuously-polydisperse $P(\sigma)$ (Eq.\ \ref{eq:sigmadist}).
To investigate this possibility, we repeated the $N = 10^5$ studies highlighted in Figs.\ \ref{fig:glasdyn}-\ref{fig:dtau}, using the 50:50 1:1.4 bidisperse $P(\sigma)$ employed in Ref.\ \cite{adhikari21} and many other studies of the glass-jamming transition \cite{liu10}.
For maximal consistency with Ref.\ \cite{adhikari21} and other previous studies, we set $\epsilon_{\rm small} = \epsilon_{\rm large} = \tilde{\epsilon}$ \cite{footepsscaling}.

\begin{figure}[h]
\includegraphics[width=2.9in]{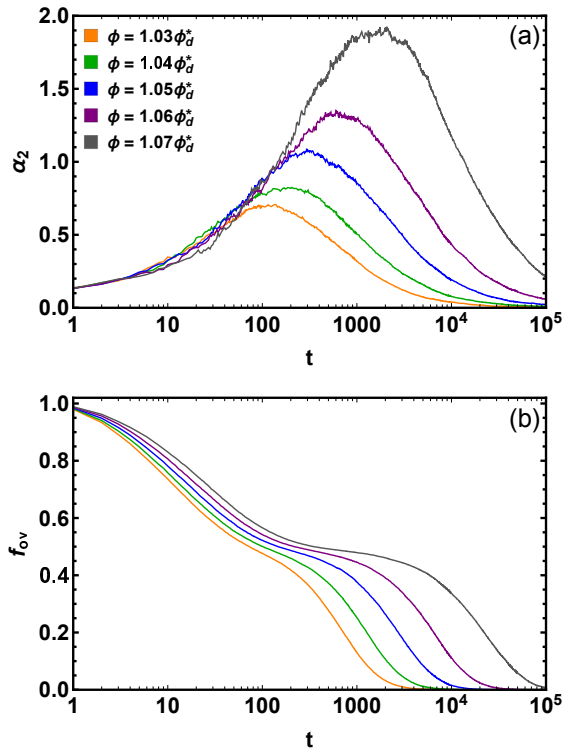}
\caption{Heterogenous dynamics of supercooled 50:50 1:1.4 bidisperse $d = 6$ liquids.  Following Refs.\ \cite{kob95,adhikari21}, we calculated $\alpha_2(t)$ and $f_{\rm ov}(t)$ separately for small and large particles.  Panel (a) shows $\alpha_2(t)$ for the large particles, while panel (b) shows $f_{\rm ov}(t)$ for all particles.}
\label{fig:bidisp}
\end{figure}

Figure \ref{fig:bidisp} illustrates two aspects of these bidisperse liquids' heterogeneous dynamics.
Panel (a) shows the large particles' $\alpha_2(t)$ for selected $\phi$.
The finite-$\alpha_{2,\infty}$ plateaus vanish, as expected \cite{kob95}.
Compared to results shown in Fig.\ \ref{fig:glasdyn}(c), the $\alpha_{\rm 2, max}$ are lower for systems with comparable $\tau^*$.
While they are substantially higher than those reported in Ref.\ \cite{adhikari21}, they are comparable to those reported in Ref.\ \cite{charbonneau12}.
For $\phi \lesssim 1.06 \phi_d^*$, \textit{if only large particles are used to estimate both $D$ and $\tau_\alpha$}, these systems have $D\tau_\alpha \sim \tau_{\alpha}^\omega$ with $\omega \simeq 0.1$, consistent with Refs.\ \cite{charbonneau12,charbonneau13b,adhikari21},

Other metrics, however, indicate that dynamics in these liquids are in fact far more heterogeneous than their $P \propto \sigma^{-d}$ counterparts, as might have been expected from their larger size asymmetry.
For example, their $f_{\rm ov}(t)$ [panel (b)] indicate a decoupling of large and small particles' dynamics that  is much stronger than that shown in Fig.\  \ref{fig:glasdyn}(b).
These data raise the question:\ which method of averaging dynamics results from polydisperse supercooled liquids best captures their essential physics?

\subsection{Non-Gaussian particle caging}
\label{subsec:NGPC}

Closely related to the above discussion is the issue of caging.
The probability $P(\vec{r},t)$ that a particle initially located at the origin is at position $\vec{r}$ at time $t$ is $P(r, t) = G_s(r,t)/A(d, r)$, where
\begin{equation}
A(d,r) =  \displaystyle\frac{d \pi^{d/2} r^{d-1} }{ \Gamma(1 + d/2)}
\label{eq:aofsigma}
\end{equation}
is the area of a $d$-dimensional hyperspherical shell of radius $r$; here we have assumed isotropy in rewriting $P(\vec{r},t)$ as $P(r,t)$.
Einstein's theory of Brownian motion predicts that $P(r,t)$ is Gaussian, and the central limit theorem \textit{requires} that it become Gaussian after sufficiently long times.
At shorter times, however, $P(r, t)$ is non-Gaussian in a very wide variety of systems, including systems near glass and jamming transitions \cite{weeks00,chaudhuri07}.
Exponential tails of form $P_E(r,t) \propto \exp[-r/\Lambda(t)]$ are universal in systems where particles have hopped a random number of times \cite{chaudhuri07,barkai20}, with $\lambda(t)$ typically growing either as $t^{1/2}$ or as $t^{1/d}$ \cite{chaudhuri07,chechkin17,barkai20,miotto21,wang09,wang12,guan14}.
In a dynamically heterogeneous liquid, these tails correspond to the high-mobility particles.  

Quantitatively predicting how $P(r,t)$ varies with $\phi$ and/or $T$ in arbitrary $d$ is an obvious goal for any theory of liquid-state dynamics.
Replica theory \cite{parisi10,charbonneau17} and dynamic DFT \cite{kirkpatrick87} assume that it is Gaussian.
MCT \cite{flenner05,schmid10}, RFOT \cite{bhattacharyya10}, the nonlinear Langevin equation theory \cite{saltzman06,saltzman08}, dynamical-facilitation-based theories \cite{berthier05b}, and CTRW-based theories \cite{chaudhuri07,chechkin17,barkai20,miotto21} all predict non-Gaussian $P(r,t)$, with varying degrees of success.
One might expect from the increase in the number of near neighbors and the simplification of liquids' local structure as $d$ increases  \cite{skoge06} that $P(r,t)$ will converge to a Gaussian form even at short times if $d$ is sufficiently large.
Ref.\  \cite{charbonneau12} showed that in fact no such convergence occurs for $t \lesssim \tau_\alpha$ over the range $3 \leq d \leq 8$, but did not examine any $t \gg \tau_\alpha$.
In light of the results presented in Figs.\ \ref{fig:glasdyn}-\ref{fig:bidisp}, it is worthwhile to examine how our larger systems' $P(r,t)$ behave in this long-time limit.

\begin{figure}[h]
\includegraphics[width=2.9in]{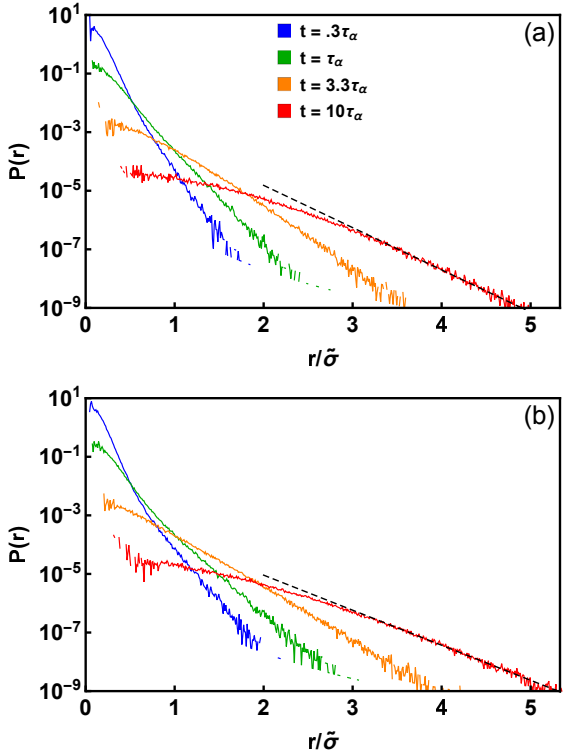}
\caption{Non-Gaussian caging of the large particles in  supercooled 50:50 1:1.4 bidisperse $d = 6$ liquids.  Panels (a-b) show results for $\phi = 1.03\phi_d^*$ and $\phi = 1.06\phi_d^*$; large particles in these systems respectively have $\tau_\alpha \simeq 7.8{\times} 10^2$ and $\tau_\alpha = 7{\times} 10^3$.  The dashed lines show fits to exponential tails with $\lambda = 0.301\tilde{\sigma}$ and $\lambda = 0.363\tilde{\sigma}$.}
\label{fig:VHC}
\end{figure}

Figure \ref{fig:VHC} shows $P(r,t)$ for the large particles in the bidisperse liquids discussed above.
Data shown in panel (b) are for a density  at which the glassy dynamics are about about ten times slower than those illustrated in panel (a).
Both panels show the same trends observed in $d = 3$  \cite{chaudhuri07,wang09,wang12,guan14}: the $P(r,t)$ are initially dominated by their exponential tails but then slowly cross over towards a Gaussian form as $t$ increases.
Two notable features are apparent.

First, contrary to what might be expected in higher-$d$ liquids but consistent with the slow decays of $\alpha_2(t)$ illustrated in Fig.\ \ref{fig:bidisp}(a), substantial exponential tails are evident even for $t = 10\tau_\alpha$.
CTRW-based theories of diffusion \cite{chechkin17,miotto21,barkai20} predict that $P(r,t)$ converges to a fully Gaussian form only after all (or nearly all) particles have hopped multiple times.
For $\phi \gtrsim \phi_d^*$, where mobility is hopping-dominated and times between hops are broadly distributed, this convergence should occur only for $t \gg \tau_\alpha$, \textit{independent of $d$.}
Second, although $P(r,t)$ for fixed $t/\tau_\alpha$ and different $\phi$ are qualitatively similar, they do not collapse.
The exponential-tail lengths $\lambda(t/\tau_\alpha)$ clearly increase faster for  $\phi = 1.06\phi_d^*$ than for $\phi = 1.03\phi_d^*$.
This result is consistent with the decoupling illustrated in Figs.\ \ref{fig:glasdyn}(b) and \ref{fig:bidisp}(b); $\lambda(t/\tau_\alpha)$ grows faster in the higher-$\phi$ system because the disparity in mobility between large and small particles is greater.

Since the slow crossovers to Gaussian $P(r,t)$ have been interpreted \cite{wang09,wang12} as a slow approach to ergodicity, suggesting that they provide a useful metric for understanding how ergodicity breaks, it would be very interesting to quantitatively compare them across multiple $d$.
While we leave this as a challenge for future work, we emphasize here that quantitative analyses of these crossovers are likely to suffer from spurious finite-size effects if the crossovers are not complete by the time the most mobile particles have traveled  distances $\gtrsim L$.
Taken together, the results shown in Figs.\ \ref{fig:dtau} and \ref{fig:VHC} suggest that avoiding such effects requires $L \gtrsim 5\sigma$, again emphasizing the need for an efficiently parallelized code like \texttt{hdMD}.

\section{Discussion and Conclusions}
\label{sec:dc}

In this paper, we described a new public-domain, open-source parallel molecular dynamics simulation package (\texttt{hdMD}) that is optimized for high spatial dimensions.
Four aspects of \texttt{hdMD}'s algorithmic implementation differ from those employed in most standard MD codes \cite{lammps,gromacs,namd,amber,hoomd,rumd}.
First, since parallelization of the force evaluations by spatial domain composition works less well in large $d$ than it does in $d \leq 3$ (owing to the larger fraction of any spatial domain that is within a distance $\tilde\sigma + s$ of its surface), \texttt{hdMD} instead employs per-atom parallelization.
Second, to further reduce interprocessor communication, \texttt{hdMD} uses a shared-memory OpenMP-based parallelization strategy rather than the more commonly employed distributed-memory MPICH-based approach.
Third, to avoid the large-for-high-$d$ memory overhead associated with storing pointers to each linked subcell's $3^d - 1$ neighboring subcells, these subcells are instead efficiently identified on the fly each time the VLs are built.
Fourth, $d$ is a parameter rather than a fixed quantity in  \texttt{hdMD}'s various subroutines, all of which have been tested for all $2 \leq d \leq 10$ and (in principle) work in arbitrary $d$.

\texttt{hDMD} is designed for maximum flexibility and extensibility.
For example, while above we presented results for a single (stiff repulsive Morse) pair potential, using a different potential or interaction cutoff radius requires only editing a few lines in \texttt{getforce()}, \texttt{getthermo()}, and \texttt{swapmove()}.
Incorporating additional diagnostics, e.g.\ calculation of the self-intermediate scattering function $F_s(q,t)$, is intended to be comparably straightforward.
For this reason, \texttt{hdMD} is written in ``plain vanilla'' C++, and software tricks like advanced vectorization techniques, SIMD or AVX intrinsics \cite{leiserson20}, and GPU offloading of the type used in several popular MD packages \cite{lammps,hoomd,anderson20}, all of which can substantially increase a code's speed but often severely reduce its legibility, have not yet implemented.
Adding any of these could substantially accelerate the code.

By examining the scalings of parallel simulation runtimes with the number of particles $N$ and the number of simulation threads $n_{\rm threads}$, we showed that three aspects of \texttt{hdMD}'s performance are already nearly optimal.
First, the runtimes scale as $t \sim N$ when $N$ is small enough for the particles' position-velocity-force (\texttt{rvf}) array to fit in the CPU's L3 cache.
Second, for large $N$, the runtimes per force evaluation increase only slowly with increasing $d$, e.g.\ by only a factor of $\sim 2$ over the range $3 \leq d \leq 6$ for simulations of $N = 10^6$ supercooled liquids at $\phi = \phi_d^*$.
Since the computational effort to rebuild all particles' Verlet lists scales as $3^d N$, this small increase is a major strength of the code.
Third, \texttt{hdMD}'s parallel efficiency is comparable to that of popular public-domain MD codes (at least for selected problems \cite{plimpton12,glaser15}), and actually increases with increasing $d$ owing to its efficiently parallelized VL-building.

The total ``size'' of each simulation described in Section \ref{sec:gd36} (as defined by the number of particles times the duration of of the simulation) was $N\mathcal{T} = 10^{10}\tau$, making them among the largest $d > 3$ supercooled-liquid simulations ever performed.
We found that dynamical heterogeneity in supercooled $d = 6$ liquids can be substantially greater than previously reported \cite{charbonneau12,charbonneau13b,adhikari21}.
In particular, we found that the $D\tau_\alpha \sim \tau_\alpha^\omega$ scaling in continuously-polydisperse systems with $\Delta = .083$ has $\omega \simeq 0.2$, which is about twice the value previously reported \cite{charbonneau13b,adhikari21} for $d = 6$.
Simulations of bidisperse systems showed $\omega \simeq 0.1$, but also demonstrated that particle caging can remain substantially non-Gaussian [as indicated by long exponential tails in particles' displacement-probability distributions 
$P(r,t)$]
for times as large as $10\tau_\alpha$.
These dynamics appear to be consistent with recently proposed, CTRW-based theories of diffusion in systems for which the exponential tails of $P(r)$ correspond to particles that have hopped a random number of times \cite{chechkin17,miotto21,barkai20}.

We also showed that the crossover to the stronger $D\tau_\alpha \sim \tau_\alpha^y$ scaling that occurs as  the continuously-polydisperse systems become sluggish \cite{charbonneau13b} occurs at a density that decreases substantially when $N$ is increased from $5000$ to $10^5$.
This decrease may arise from larger systems' ability to accommodate larger cooperatively rearranging regions (CRRs).
Specifically, our results are consistent with the hypothesis that for $\phi  > \phi_c(N)$, $\tau_\alpha$ grows faster with $\phi$ in larger systems of size $N' > N$ than in smaller systems of size $N'' < N$ because the former can accommodate larger CRRs which have a correspondingly larger $\tau_\alpha$ \cite{starr13}.
While the validity of this hypothesis can depend on both temperature and the model employed \cite{berthier12}, our results nonetheless suggest that the conclusions of many previous studies of supercooled liquids in $d > 3$ which employed fixed $N < 10^4$ and $L$ that decrease as $N^{-1/d}$ (e.g.\ Refs.\ \cite{lue06,skoge06,vanMeel09b,charbonneau11,charbonneau12,charbonneau13b,charbonneau14c,charbonneau14,adhikari21,berthier19c,berthier20,morse21}) may have been substantially influenced -- at least in their quantitative details -- by finite-size effects.
We have demonstrated that \texttt{hdMD} is well-suited to determining whether this is so.

Finally we emphasize that \texttt{hdMD} is also well-suited to studying open problems that are less directly related to the glass-jamming transition.
For example, studies of melting dynamics across multiple $d$ can shed light on how melting is affected by the symmetries of the crystal lattice and by decorrelation \cite{skoge06} of the liquid state.
Previous  studies of melting in $d > 4$ (e.g.\ \cite{lue10,estrada11,lue21,charbonneau21}) have all employed $N < 6{\times} 10^4$, and most have employed much smaller systems; this has severely limited the accessible size range of any crystal-fluid interfaces.
We have used \texttt{hdMD} to simulate the (nonequilbrium) melting of an $N = 6.25\cdot 10^5$-atom $E_7$ crystal (the densest lattice in $d = 7$) subjected to a temperature ramp at constant pressure and will report our results elsewhere.

The \texttt{hdMD} source code is publicly available and can be downloaded from our group website (http://labs.cas.usf.edu/softmattertheory/hdMD.html).

We are grateful to Patrick Charbonneau for numerous helpful discussions.
This material is based upon work supported by the National Science Foundation under Grant No.\ DMR-2026271.


\begin{thebibliography}{93}
\bibitem [{foo({\natexlab{a}})}]{footepsscaling}%
  \BibitemOpen
  \href@noop {} {} \bibinfo {note} {The more common
  $\sigma_i$-independent $\epsilon_i$ \cite{kob95} can be implemented by
  editing a single line in \texttt{initRMSE()}.}\BibitemShut {Stop}%
\bibitem [{foo({\natexlab{b}})}]{footnandi}%
  \BibitemOpen
  \href@noop {} {}  \bibinfo {note} {This value $x \simeq
  0.3$ is not universal; for example, Ref.\ \cite{nandi21} found $x \simeq
  0.4$.}\BibitemShut {Stop}%
\bibitem [{foo({\natexlab{c}})}]{footgarrahan}%
  \BibitemOpen
  \href@noop {} {} \bibinfo {note} {Indeed, Ref.\
  \cite{kim17} implied that this effect may have been the origin of Ref.\
  \cite{charbonneau13b}'s conclusion that $\omega \to 0$ as $d \to
  8$.}\BibitemShut {Stop}%
\end{thebibliography}

%

\end{document}